\documentclass[12pt]{spieman}  % 12pt font required by SPIE;
\usepackage{amsmath,amsfonts,amssymb}
\usepackage{graphicx}
\usepackage{setspace}
\usepackage{tocloft}
\usepackage{multirow}
\usepackage{xcolor}
\title{Mitigating the Non-Linearities in a Pyramid Wavefront Sensor}
\author[a,b]{Finn Archinuk}
\author[c]{Rehan Hafeez}
\author[a,b]{Sébastien Fabbro}
\author[a,b]{Hossen Teimoorinia}
\author[a,b,*]{Jean-Pierre Véran}

\affil[a]{National Research Council Canada, 5071 West Saanich Road, Victoria, Canada, V9E 2E7}
\affil[b]{University of Victoria, 3800 Finnerty Road, Victoria, Canada, V80 5C2}
\affil[c]{University of British Columbia, 2329 West Mall, Vancouver, Canada, V6T 1Z4}

\cftpagenumbersoff{figure}
\cftpagenumbersoff{table} 
\begin{document} 
\maketitle

\begin{abstract}

For natural guide star adaptive optics (AO) systems, pyramid wavefront sensors (PWFSs) can provide \textcolor{black}{a} significant increase in sensitivity over the traditional Shack-Hartmann, but at the cost of a reduced linear range. When using a linear reconstructor, non-linearities result in wavefront estimation errors, which can have a significant impact on the image quality delivered by the AO system. 
Here we simulate a wavefront passing through a PWFS under varying observing conditions to explore the possibility of using a non-linear machine learning model to estimate wavefront errors \textcolor{black}{and compare with a} linear reconstruction.
We find significant \textcolor{black}{potential} improvement\textcolor{black}{s in delivered image quality} even with \textcolor{black}{computationally simple} models, underscoring the need for further investigation of this approach.
\end{abstract}

% Include a list of up to six keywords after the abstract
\keywords{adaptive optics, machine learning, wavefront reconstruction, astronomy}

% Include email contact information for corresponding author
{\noindent \footnotesize\textbf{*}Jean-Pierre Véran,  \linkable{Jean-Pierre.Veran@nrc-cnrc.gc.ca} }

\begin{spacing}{2}   % use double spacing for rest of manuscript

\section{Introduction}
\label{sec:intro}
\textcolor{black}{When observing the sky at visible or infrared wavelengths with a large ground-based telescope, atmospheric turbulence causes random distortions in the incoming light (i.e. wavefront errors) that significantly reduce the resolution and contrast of the recorded images. These effects can be corrected in real-time with an adaptive optics (AO) system, and most present and future optical telescopes include one or several AO systems. While AO systems cannot provide a perfect correction, the residual wavefront errors are small enough so that the delivered image quality becomes limited by the diffraction of the telescope, a fundamental limit. This results in higher resolution, higher contrast astronomical images unveiling finer details of the structures in the universe. AO also dramatically increases the sensitivity of the observations, significantly reducing the required exposure time to reach a given signal-to-noise ratio on scientific targets, therefore allowing more targets to be observed each night.}

\textcolor{black}{A basic AO system, also called single-conjugate AO (SCAO) system, consists of three major components: a deformable mirror (DM), which corrects the wavefront distortions using actuators pushing and pulling a reflective surface; a wavefront sensor (WFS), which measures the residual wavefront errors on a bright guide star; and a real-time controller (RTC), which estimates the wavefront from the WFS measurements and updates the DM commands so that the residual wavefront errors are minimized. An AO system, therefore, is a closed-loop feedback system \textcolor{black}{that} needs to be operated at typical frame rates of $\sim$1 kHz in order to keep up with changes in atmospheric turbulence. This means that every millisecond, a set of WFS measurements is obtained, and the DM shape is updated. Beyond SCAO, more sophisticated AO systems involving several WFSs and possibly several DMs have been developed in order to increase the size of the corrected field. However, SCAO systems are very relevant, especially when working on scientific targets close to a bright star, which can be used as a guide star for the WFS. This is the case for the so-called extreme AO systems, such as the Gemini Planet Imager (GPI)\cite{Macintosh2018}, which aim at producing very high contrast images in which faint stellar companions, such as exo-planets, can be found.}

\textcolor{black}{The performance of the AO system is almost always limited by the WFS's ability to accurately measure the instantaneous residual wavefront from a limited number of photons, making the WFS an absolutely critical component. The traditional Shack-Hartmann WFS has been widely used in existing AO systems because it provides a high-level of linearity in reconstructing the wavefront from the WFS. The reconstructor is then implemented as a simple matrix-vector multiplication (MVM), an easily parallelizable process that can be executed with very low latency on modern computers \cite{chew2006}. However, different WFSs such as the Pyramid WFS \cite{ragazzoni2018} (PWFS) have been introduced recently because they are more sensitive, providing a more accurate measurement for a given light level, or conversely, providing the same accuracy for a lower light level (i.e. a fainter guide star). This increase in sensitivity, however, comes with a loss of linearity, which creates additional errors in the wavefront reconstruction process, when a linear reconstructor is used \cite{shatokhina2020}. With a PWFS, the trade-off between increased sensitivity and loss of linearity can be adjusted by modulating the image of the guide star around the tip of the pyramid during light integration on the WFS detector, and by adjusting the modulation radius\cite{Verinaud2004}. However, even when this trade-off is optimized, the non-linearity errors can be quite significant. For example, NFIRAOS\cite{Crane2018}, the first light AO system for the future Thirty Meter Telescope, has a PWFS for natural guide star observations, and the PWFS non-linearity effects account for 64 nm of RMS wavefront error, out of a total budget of 156 nm RMS on a magnitude 8 natural guide star \cite{hardy2017}. Using the so-called Marechal approximation\cite{Mahajan1983}, RMS wavefront errors can be directly translated into the Strehl ratio, a quantitative metric for image quality. 156 nm RMS corresponds to a Strehl ratio of 70.2\% at a wavelength of 1.65 {\textmu}m (H-Band), down from 74.6\% with no non-linearity errors.}

In this paper, we propose to evaluate \textcolor{black}{how one can mitigate non-linearities in a PWFS by implementing a non-linear wavefront reconstructor derived using deep learning. We limit ourselves to a traditional AO setup where each WFS measurement is processed for wavefront reconstruction independently, therefore focusing on spatial effects as opposed to temporal effects. We refer to our previous work \cite{Hafeez2022} for an attempt to use machine learning in order to predict atmospheric turbulence by taking advantage of short-range temporal correlations. Specifically,} a convolutional neural network (CNN) as a reconstructor \textcolor{black}{is evaluated as a substitute for a linear system}. The CNN is built through standard machine learning methods by training on simulated wavefront maps representative of typical wavefronts measured by AO systems. We show that the CNN is able to capture non-linearities in the measurement process and provide a reconstruction accuracy that is significantly better than the linear reconstructor.

\section{Related Work}

\textcolor{black}{M}odern observatories \textcolor{black}{are increasingly employing PWFSs,} including Keck\cite{IRPWFSatKeck}, \textcolor{black}{the} Large Binocular Telescope \cite{PWFSatLargeBT}, \textcolor{black}{the} Large Magellan Tele\textcolor{black}{s}cope \cite{NGSatGMT} and \textcolor{black}{the} Thirty Meter Telescope \cite{Vran2015PyramidVS} \textcolor{black}{currently under construction}. The \textcolor{black}{PWFS} tend\textcolor{black}{s} to replace \textcolor{black}{the} more traditional Shack-Hartmann WFS in \textcolor{black}{N}atural \textcolor{black}{G}uide \textcolor{black}{S}tar AO systems because of \textcolor{black}{its} increased sensitivity which enables \textcolor{black}{improved AO correction and/or} larger sky coverage.

\textcolor{black}{The sensitivity of the PWFS (i.e. optical gain) changes depending on the level of correction provided by the AO system (better AO correction = lower residuals = higher sensitivity), which in turn depends on observing conditions. If left uncompensated, these changes in sensitivity would cause the AO loop gain to fluctuate, potentially leading to under-performance (gain too low) or even instabilities (gain too high), as well as errors in compensating for non-common path aberrations (aberrations seen by the science channel but not seen by the WFS channel or vice-versa)\cite{Esposito2015}. This problem has been mitigated by estimating the optical gain in real time as conditions change and applying its inverse as part of the wavefront reconstruction process\cite{Esposito2020}. It was then recognized that each correction mode had its own optical gain\cite{Deo2018}, and that adjusting the modal optical gains can not only account for changes in observing conditions but if the adjustment can be performed often enough, mitigate non-linearities as well\cite{Chambouleyron2020}\cite{Chambouleyron2021}. Weinberger et al.\cite{Weinberger2022} have even proposed to use neural networks to estimate the optical gains. However, all these methods assume that the modal optical gains are independent, which is only an approximation, especially in very non-linear conditions, such as in the case of an unmodulated PWFS. So very recently, this idea was extended with the SIMPC method\cite{Agapito2023}, which computes the entire wavefront reconstruction matrix around the expected level of wavefront residual, in effect linearizing the wavefront reconstruction problem around typical AO residual levels, as opposed to no residuals in the traditional method. How to implement this approach in practice, however, remains a topic for research, as the reconstruction matrix would still have to be updated when observing conditions, and therefore AO residuals, change. In this paper, we explore the alternative approach of giving up on the linear reconstructor altogether and replacing it with a non-linear reconstructor implemented in the form of a CNN.}

\textcolor{black}{Neural network approaches have been proposed in order to improve AO correction, but most of this work has been focused on trying to predict the atmospheric turbulence in order to reduce the lag error inherent to all AO systems \cite{Hafeez2022} \cite{Guyon2017} \cite{Wong2022NN}}. Machine learning techniques have recently been recognized \textcolor{black}{for their potential} to mitigate non-linearities intri\textcolor{black}{n}sic to WFSs \cite{Nishizaki2019}. Some advances have been made to apply these techniques to the Shack-Hartmann WFS\cite{He2021} and to design better WFSs\cite{Vera:2021}. A recent paper provides an overview on machine learning applications for wavefront sensing\cite{Wong2022}, but only identifies a single application to the PWFS, in ophthalmology\cite{Diez2008}, where the operating conditions are quite different from those in astronomy. So far, machine learning techniques have been used for highly non-linear wavefront sensing problems. For example,
Xivry et al\cite{Xivry2021FocalPW} looked at measuring non-common path aberrations directly in the science focal plane (\textcolor{black}{i.e.} focal plane WFSing)\textcolor{black}{. Landman et al. \cite{Landman_2020} demonstrated that CNNs can be used to aid the reconstruction of wavefronts measured with a WFS purposefully made non-linear in order to increase its sensitivity}. While the task and constraints of their problems differ from what we focus on, their use of deep convolutional architectures demonstrate\textcolor{black}{s} the adoption of machine learning methods for real-world non-linear problems. 

\textcolor{black}{Concurrently, advancements in image-based machine learning have surged.} Transformer architectures have been adapted for images \cite{vit} \cite{swin_transformers}; however, \textcolor{black}{they still tend to require more parameters and data to train with, as a result they are} too computationally expensive for millisecond rate real-time implementation.
Residual networks (ResNets)\cite{resnet} form the backbone of many image-based models. Their efficacy is due to residual connections, which allow more layers to be trained. The advances found by vision transformers \textcolor{black}{have been} analyzed and applied to ResNets, which led to \textcolor{black}{the development of} ConvNeXt \cite{convnext}, \textcolor{black}{which further builds upon ResNets by parallelizing computational pathways}. \textcolor{black}{These promising advancements in computer vision hold potential for our case. However, achieving sub-millisecond inference times imposes constraints on our choice of architectures, limiting flexibility, as explored in Section \ref{sec:computational limitations}.}

\section{Data Source}
\label{sec:data source}

\textcolor{black}{In this study, we utilize simulated random wavefront maps that are generated using the power spectrum method. This method entails taking the Fourier transform of the square root of the desired power spectrum, with the addition of a random phase at each frequency $f$.} \textcolor{black}{Most AO systems have a closed-loop architecture, which means that the WFS does not see the full atmospheric turbulence but instead measures the correction residuals. Accordingly, we have simulated} wavefront maps intended to represent typical AO residual wavefronts, which are sensed by the PWFS. Their power spectrum \textcolor{black}{follows} a $f^{-2}$ power law, which is shallower than the Kolmogorov $f^{-11/3}$ power law representing the uncorrected atmospheric turbulence. An $f^{-2}$ power law is typical for AO corrected wavefronts for spatial frequencies within the correction range of the DM\cite{Jolissaint2010}. However, in order to include more diversity in our training set, as well as to test the robustness of the model, we have also generated wavefront maps with $f^{-1.8}$ and $f^{-2.2}$ power laws. When generating data with power law $f^{-p}$, we refer to $p$ \textcolor{black}{as} the \textit{f-value}.

The wavefront maps are generated for a Gemini-like D=8m telescope on a square grid of 176x176 pixels, which provides a sampling of 22 pixels per meter \textcolor{black}{when projected on the primary mirror of the telescope}. The pupil is circular with a central obscuration of 1.0m. For this pupil, the Karhunen—Lo\`eve  (KL) modes of the Kolmogorov turbulence have been computed, and each wavefront map is mathematically projected onto the first 1603 KL modes. The resulting 1603 coefficients are the ``true" modal coefficients, which, if applied to a modal DM would minimize the residuals. Of course, in a real system, the true coefficients are not available and must be estimated from the WFS measurements. The goal, then is to minimize the error between the estimated coefficients and the true coefficients in order to maximize the delivered image quality.

\begin{figure}[h]
\begin{center}
\begin{tabular}{c}
\includegraphics[width=16cm]{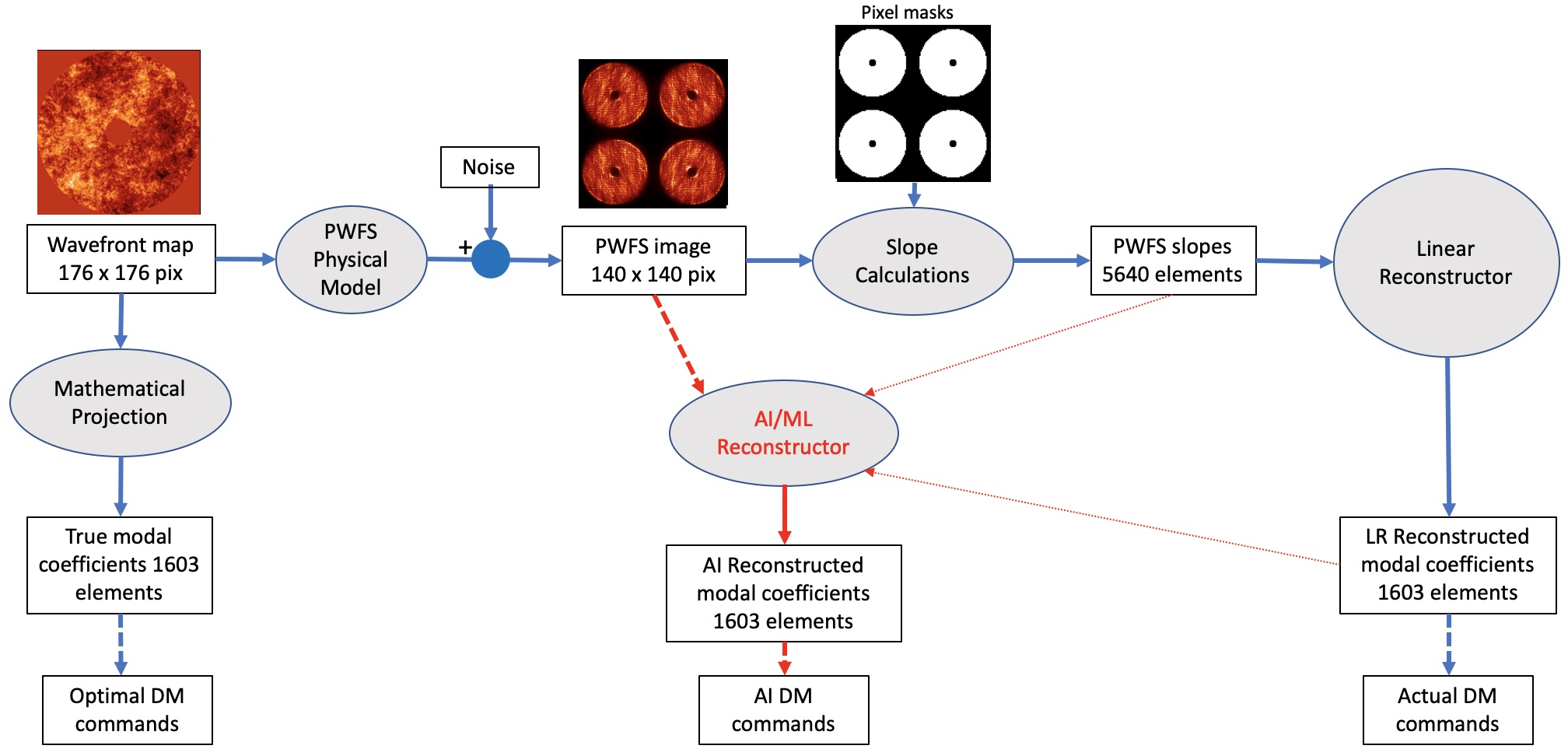}
\end{tabular}
\end{center}
\caption 
{ \label{fig:sim_pipeline}
Outline of the simulation pipeline. \textcolor{black}{Thin red dotted lines are possible inputs to the ML model that we did not pursue in this work. The thick red lines show the inputs and outputs of the proposed ML model.}}
\end{figure}

The PWFS is simulated using the physical optics PWFS module included in PASSATA \cite{agapito2016}. The PWFS is set up so that it mimics the Gemini P\textcolor{black}{la}net Imager (GPI) 2.0 PWFS, which uses an EMCCD220 with 60 pixels in the diameter of each of the four pupil images\cite{Fitzsimmons2020}. The PWFS is modeled to be sensitive to a 300 nm wide band centered on $\lambda=750$nm. The flux available to the PWFS is derived from the magnitude value of the target, assuming an A0 star, and affects the photon noise applied to the PWFS images. \textcolor{black}{Noiseless} simulations, corresponding to the case of a very bright guide star \textcolor{black}{are also} performed for reference purpose\textcolor{black}{s}. Each PWFS image is captured on a 140x140 pixel grid. All the measurements are obtained with a $3\lambda/D$ modulation, which is typical for such systems\cite{Madurowicz2020}. The software determines the correct masks to extract the four pupil images, from which 5640 X and Y slopes can be computed. A linear reconstructor is used to obtain an estimate of the 1603 modal coefficient\textcolor{black}{s from the slope vector} for each wavefront map. The linear reconstruction is obtained via the \textcolor{black}{singular value decomposition} inversion of a modal interaction matrix, which contains the measured slopes of each KL mode. This interaction matrix and its inverse are obtained directly from the PASSATA software. \textcolor{black}{The software also} ensures that the modes\textcolor{black}{, when presented for the interaction matrix acquisition, } have the proper amplitude to stay within the linear range of the PWFS. %when presented for \textcolor{red}{the} interaction matrix acquisition,
Figure \ref{fig:sim_pipeline} outlines this simulation process graphically. 

The metric to be minimized is the quadratic norm \textcolor{black}{- i.e. the root sum square (RSS) - } of the residual coefficients, which \textcolor{black}{is} the difference between the 1603 reconstructed and the true modal coefficients. Since the KL modes are orthonormal over the circular pupil function, minimizing the RSS of the residual coeffic\textcolor{black}{i}ents maximizes the Strehl ratio - i.e. the optical quality - of the delivered image. The RSS of the residual coefficients is directly related to the Root Mean Square \textcolor{black}{(RMS)} of the residual wavefront error\textcolor{black}{. However, } the latter also includes the wavefront errors of a higher order than the \textcolor{black}{first }1603 KL modes, which cannot be corrected by the system (fitting error).  

It is well known that when the RMS wavefront error is low, the linear reconstructor performs well, but as the RMS increases, the measurement becomes less linear and the reconstruction error increase\textcolor{black}{s}\cite{shatokhina2020}. Current PWFS-based AO system accepts this trade-off \cite{IRPWFSatKeck}\cite{PWFSatLargeBT}\cite{NGSatGMT}\cite{Vran2015PyramidVS}. With the limitations of a linear reconstructor stated, we use the linear model as a baseline for quality.

Data is generated in groups of 10,000 frames characterized by \textcolor{black}{the} magnitude of their guide star and the \textit{f}-value. Generated wavefront maps are intentionally designed to be statistically independent in order to maximize the information for training the \textcolor{black}{C}NN. %\textit{f}-values were selected to be between $f^{-1.8}$ and $f^{-2.2}$, which, as discussed above, is typical for residual wavefronts in a closed loop AO system.
\textcolor{black}{The m}odels we evaluate are trained on magnitude 8 or 9 simulated sources and tested on magnitude 8, 9, and 10 data.

In order to obtain data sets representing a variety of conditions, we scale each wavefront map to have an RMS wavefront error between 0 and 200 nm. This spans a reasonable range of residual wavefront amplitudes, which \textcolor{black}{vary} depending on observing conditions \textcolor{black}{and parameters of the AO system under study}. For each frame, the RSS value of the 1603 coefficients \textcolor{black}{will be slightly below the RMS wavefront error} because of the fitting error, as discussed previously. 

\begin{figure}[h]
\begin{center}
\begin{tabular}{c}
\includegraphics[width=16cm]{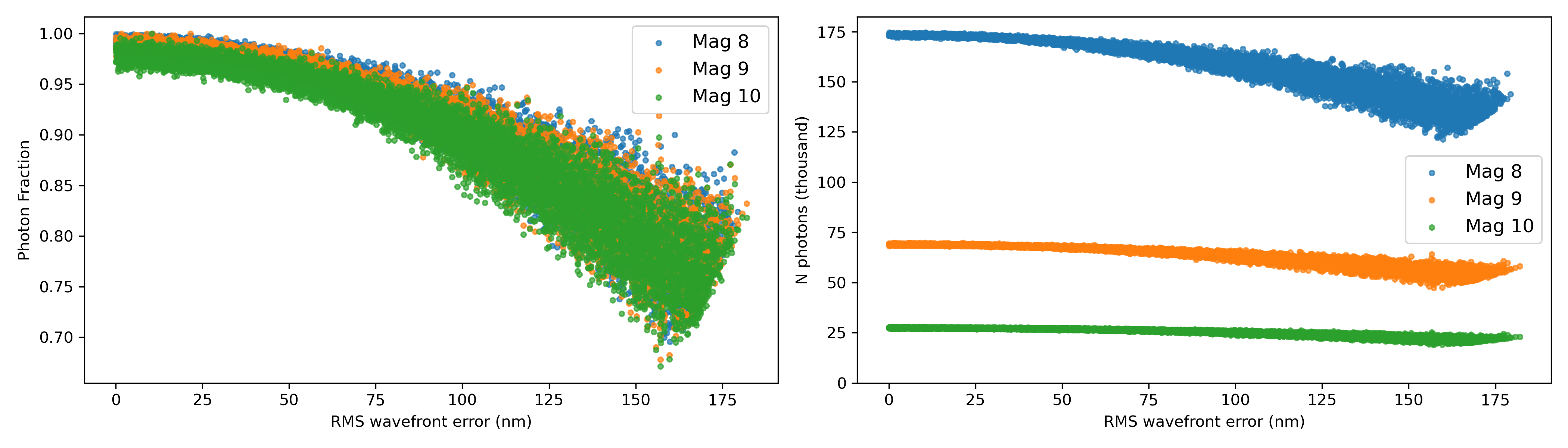}
\end{tabular}
\end{center}
\caption 
{ \label{fig:photon_loss}
Flux in the geometric pupil images as a function of the wavefront amplitude for p=2 and various magnitudes. Left: expressed as a fraction of the total \textcolor{black}{number of photons hitting the pupil. This is normalized to the sample with the maximum number of photons for each magnitude. R}ight: expressed in absolute number of \textcolor{black}{photons}.}
\end{figure}

Finally, we explore the fraction of the flux actually contained in the four pupil images used to compute the slopes. Because of diffraction, some photons land outside the geometric pupil images. Figure \ref{fig:photon_loss} shows that as the wavefront amplitude increases, a larger fraction of the flux is diffracted outside the geometric pupil images. \textcolor{black}{Diffraction effects result in the loss of up to 30\% of photons not reaching the geometric pupil images for frames with the highest RMS values.} \textcolor{black}{These photons are unused in a traditional linear reconstructor where slopes are first computed from pixels within the geometric pupil images, but we suspect that they might be key to modeling the non-linearities of the PWFS.}

\section{Data Formatting and Preprocessing}
\label{sec:data formatting}

The two potential inputs for our model are CCD frames or slopes. Slopes are a reduced representation of the wavefront, which is beneficial for simplifying the \textcolor{black}{C}NN model by reducing the number of input features. However, there is information loss in this reduction process, confirmed by \textcolor{black}{our own} experiments, which showed better accuracy when inputs from the original CCD frames are used. \textcolor{black}{This could be attributed, at least in part, to the flux beyond the boundaries of the geometric pupil images, which is disregarded during the slope computations, as previously discussed.}. 

\textcolor{black}{The initial processing step involves converting the WFS frame into a reduced-intensity image by normalizing it with the total count and then subtracting the normalized WFS image corresponding to a flat wavefront}: 
\[\Delta I (\phi) = \frac{I(\phi)}{\Sigma I(\phi)} - \frac{I(\phi = 0)}{\Sigma I(\phi = 0)}\]

\noindent where $\phi$ is the wavefront map, $\Delta I (\phi)$ is the reduced intensity image, $I(\phi)/\Sigma I(\phi)$ is the normalized CCD frame and $I (\phi = 0)/\Sigma I(\phi = 0)$ is the normalized WFS image for a flat wavefront. \textcolor{black}{The utilization of reduced intensity images offers the advantage of robustness to variations in total illumination, as well as having a flat wavefront image as the reference with zero intensity.} Using reduced intensities instead of computing slopes has become a standard practice as the former contains more information than the latter for the wavefront reconstruction process\cite{Chambouleyron2020}. Of course, for real non-simulated images, standard image \textcolor{black}{pre}processing, including flat field and background removal, would also have to be performed.

After creating the reduced intensity image, the floating point values of the frame \textcolor{black}{require} scaling. Raw values \textcolor{black}{are} close to zero, and neural networks perform best when data points utilize the space between ±1.
We tried a variety of methods to return these values to a useful range: frame-wise scaling, where the frame has a mean of zero and standard deviation of one; pixel-wise scaling, where each pixel location in the training data is scaled to have \textcolor{black}{a} mean \textcolor{black}{of} zero and standard deviation \textcolor{black}{of one}; and fixed scaling, where the whole frame is multiplied by a fixed constant.

Frame-wise and fixed scaling are the most obvious scaling methods, relying only on the information within the frame.
Fixed scaling maintains a direct connection to the reduced intensity image by upscaling the resulting frame by a constant factor. Our experiments suggest that this scaling method was the most robust for changes in magnitude for the set of sources we evaluated. An empirical value of 1000 works well and could be considered as a hyperparameter when further optimizing this model\textcolor{black}{, or for adapting this model to a PWFS with a different number of illuminated pixels}.

\begin{figure}
\begin{center}
\begin{tabular}{c}
\includegraphics[width=16cm]{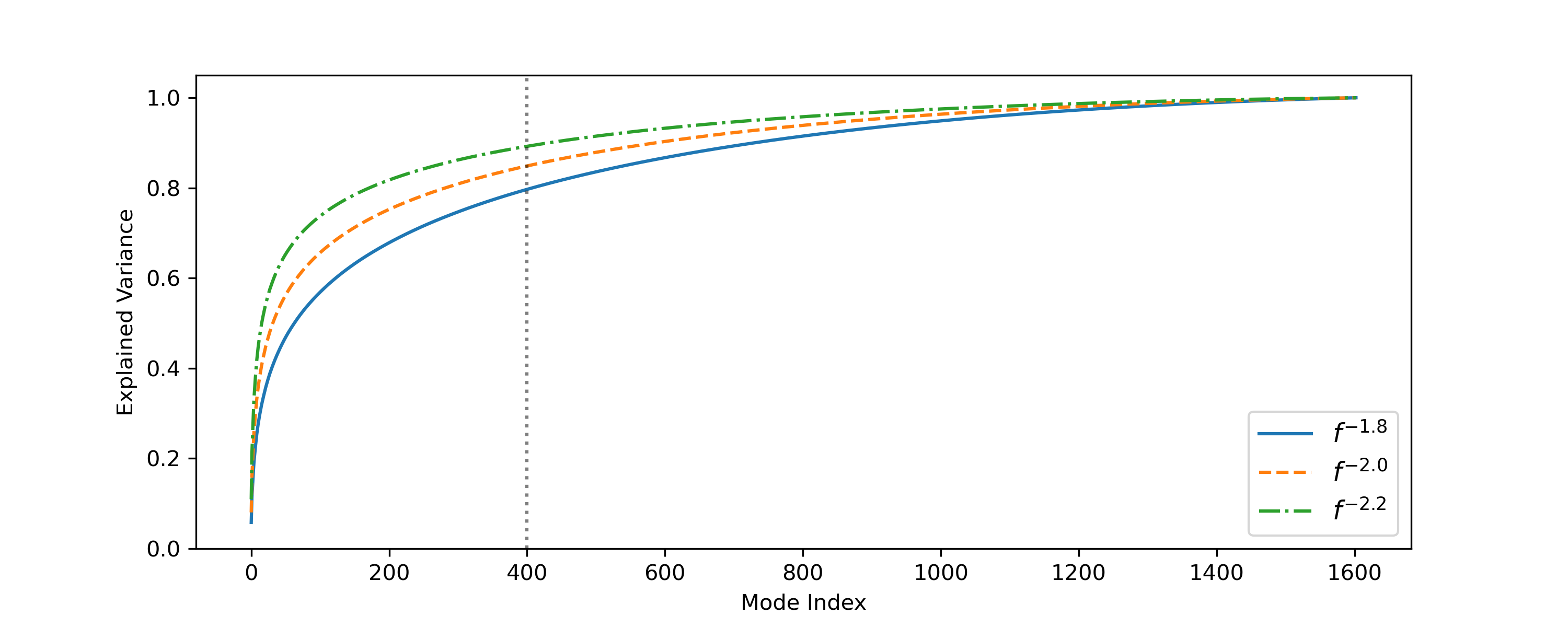}
\end{tabular}
\end{center}
\caption 
{ \label{fig:variance_by_fraction}
\textcolor{black}{The e}nergy of cumulative modes. Each line represents 10,000 modes with the specified f-value.}
\end{figure}

Figure \ref{fig:variance_by_fraction} observes the fraction of \textcolor{black}{total modal RMS wavefront error} given the number of modes considered. This fraction depends on the \textit{f}-value, and the three \textit{f}-values we evaluate are plotted. As expected, low order modes carry more energy when a steeper (higher \textit{f}-value) power-law is used. In our work, we focus on the reconstruction of the first 400 modes which\textcolor{black}{, as shown in the figure,} correspond to at least 75\% of the total wavefront RMS.

\section{Neural Network Architecture}
\label{sec:computational limitations}

Computer vision research has mostly been focused on deep neural network architectures \textcolor{black}{over} the last ten years. One of the most adopted deep architecture\textcolor{black}{s} for supervised learning tasks has been the Convolutional Neural Network (CNN), which exploits translational symmetry by applying successive learned filters in a hierarchical manner\textcolor{black}{, that is, CNNs allow us to efficiently find spatial pattern across the input. We performed significant architecture and hyperparameter tuning, with the greatest impacts on wavefront reconstruction quality being informed by the number and size of filters, and the number of convolutional layers.} Here \textcolor{black}{we present} a CNN architecture that performs well \textcolor{black}{across} the data sets we evaluated.

\begin{figure}
\begin{center}
\begin{tabular}{c}
\includegraphics[width=16cm]{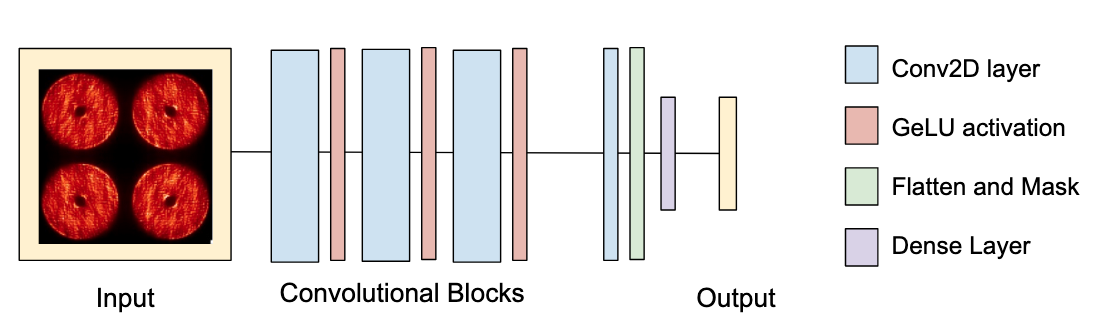}
\end{tabular}
\end{center}
\caption 
{ \label{fig:cnn architecture}
\textcolor{black}{Our proposed architecture has three convolutional blocks, followed by flattening and modal reconstruction. The input is a pupil image sample following the reduced intensity preprocessing and the output are the first 400 modes.}}
\end{figure} 

\textcolor{black}{A visual representation of our CNN architecture is found in Figure \ref{fig:cnn architecture}. The model} uses 3 convolutional \textcolor{black}{blocks}, each with 16 (5x5) filters. A \textcolor{black}{fourth} convolutional layer projects these to a single \textcolor{black}{channel}. We implemented a custom layer to apply the pupil mask \textcolor{black}{to} remove the background pixels \textcolor{black}{from the CCD frame. The \textcolor{black}{reconstructed modal coefficients} - the output of the model - are a linear combination of the remaining pupil pixels without a bias coefficient. The custom layer leverages AO-specific knowledge about the task to remove a considerable number of model parameters ($\approx$ 1 million) without significant impact on the model quality. Removing the bias term from the final fully connected layer has a minimal impact \textcolor{black}{o}n the model quality, but this choice is informed by the fact that the modes from the reduced intensity input should be centered at zero}. This architecture has approximately 5.1 million trainable parameters.

Optimizing the architecture for real-time inference would require considerable work and is outside the scope of this analysis. Larger models have been demonstrated to run at 6 milliseconds on nVidia V100 GPUs
\cite{Thomas2021ImprovingAI} suggesting our model is likely to be able to be optimized to the required \textcolor{black}{latency ($<\sim$1ms)} on more modern GPUs.

\subsection{\textcolor{black}{Model Training}}
\label{sec:model training}
\textcolor{black}{M}odels were trained with 17,100 \textcolor{black}{simulated} frames and stopped when the \textcolor{black}{error of the} validation \textcolor{black}{set of 900 frames} stopped \textcolor{black}{decreasing}. Models were trained on either magnitude 8 or 9 data. We used \textcolor{black}{the} AdamW \cite{adamW} \textcolor{black}{optimizer} with a learning rate of 8e-5 and weight decay of 1e-5. Training used a batch size of 256. Model quality was assessed against the ground truth modes \textcolor{black}{using mean squared error as a metric. Our intention is to optimize the entire \textcolor{black}{set of reconstructed modal coefficients} instead of specific modes; therefore, we did not adjust the scaling of the modal coefficients during training. We let larger magnitudes of the lower modes act as output weighting so that the lower modes (especially tip, tilt, and defocus) were proportionally more important for the model to optimize for}.
By separating \textcolor{black}{reconstruction} quality by total RMS wavefront error and guide star magnitude, we gain a better understanding of the \textcolor{black}{reconstruction} limitations of the models. From these metrics, we were able to develop model architectures and regularization hyperparameters iteratively.
We observed our model performing poorly in the low RMS region relative to the linear \textcolor{black}{reconstructor} and attempted sample weighting so that low RMS frames were more impactful when calculating the gradient \textcolor{black}{to update model parameters}. \textcolor{black}{S}ample weighting followed the exponential decay equation: $ab ^ x$, where $a$ was set to an initial value of 1000, $b$ was the exponential decay of 0.975 and $x$ was equal to \textcolor{black}{the} RMS of the sample.

\begin{figure}
\begin{center}
\begin{tabular}{c}
\includegraphics[width=16cm]{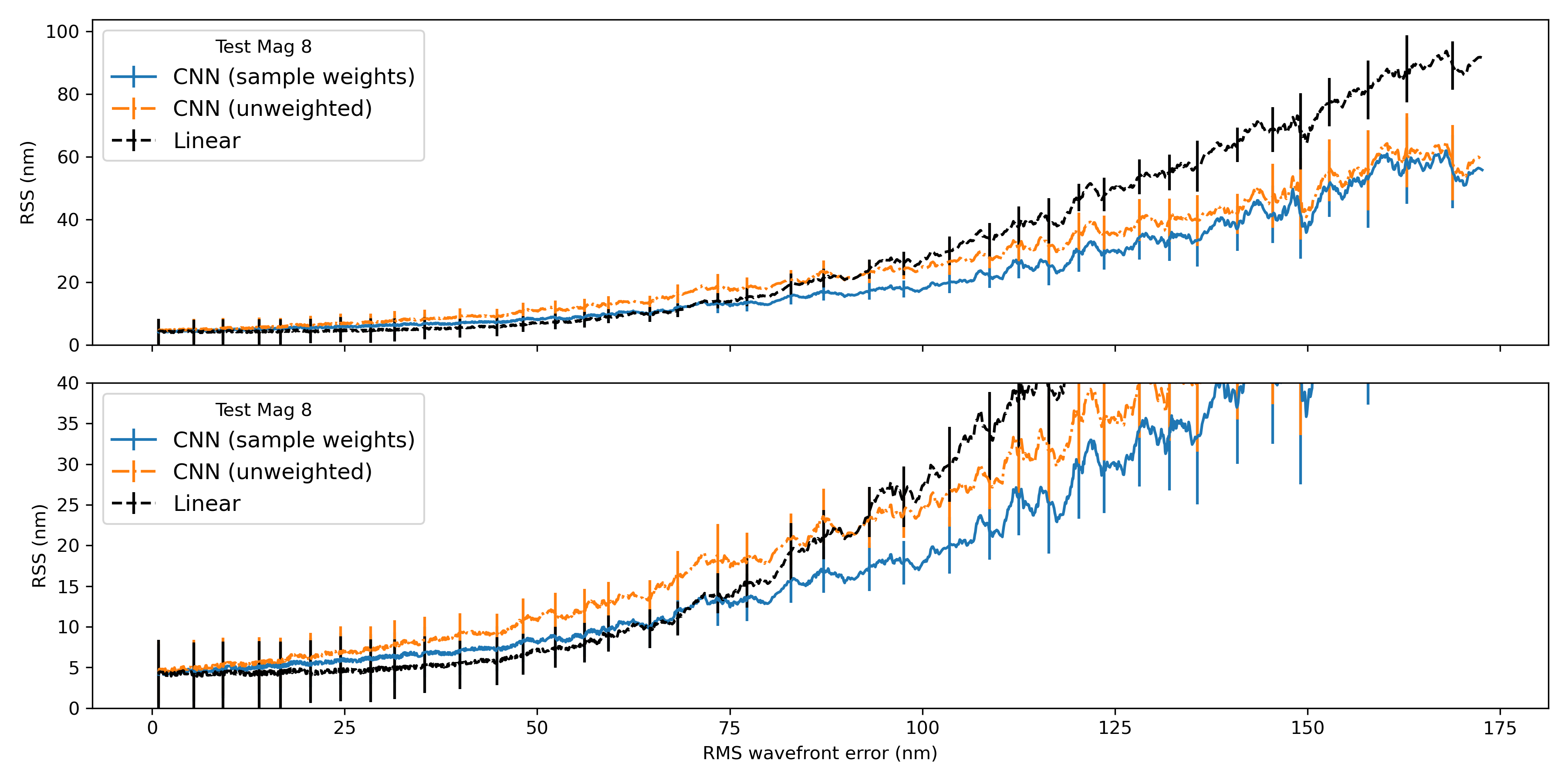}
\end{tabular}
\end{center}
\caption 
{ \label{fig:weight regularization}
\textcolor{black}{Weighting samples by RMS value improves model performance.
The top panel shows the overall change, and the lower panel shows a zoomed-in view of the lower ($<$ 125nm) RMS cases. Samples were weighted based on their RMS, with low RMS samples being higher weighted. Moving averages are plotted to help visualize the trend, and error bars are the standard deviation of the window.}}
\end{figure}

\textcolor{black}{Figure \ref{fig:weight regularization} shows the effects of sample weighting on otherwise identical models. Results are the RSS of the reconstructed modal coefficients vs ground truth and have been separated by RMS of the input wavefront error. The top panel shows improvement at the higher RMS wavefront error levels, and the lower panel zooms into the lower RMS wavefront error cases}. We expected sample weighting to cause a loss in \textcolor{black}{reconstruction} quality for the higher RMS region; however, it acted as a regularization term and improved the \textcolor{black}{reconstruction} quality over all RMS values. \textcolor{black}{The reason this may be acting as a regularizer is that by forcing the model to focus on the linear region, simpler solutions were imposed across the model.}

\section{Results}
\label{sec:results}

\subsection{\textcolor{black}{Wavefront reconstruction}}
\begin{figure}
\begin{center}
\begin{tabular}{c}
\includegraphics[width=16cm]{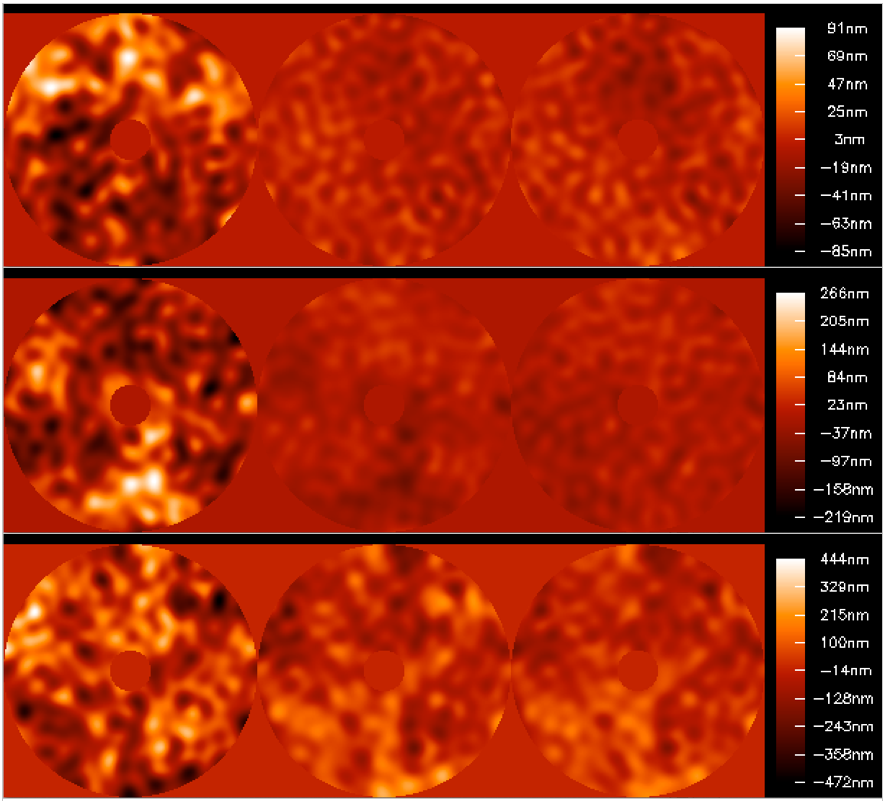}
\end{tabular}
\end{center}
\caption 
{ \label{fig:reconstructed wavefronts}
\textcolor{black}{Left: Incoming wavefront map, 400 KL modes (30/80/134 nm RMS). Middle: residual wavefront error map after linear reconstruction (8/25/85 nm RMS). Right: residual wavefront error map after CNN reconstruction (10/19/69 nm RMS).}}
\end{figure}

\textcolor{black}{Figure \ref{fig:reconstructed wavefronts} shows sample reconstruction residuals, comparing the traditional linear reconstruction method and the CNN approach. We have limited the wavefront to the first 400 KL modes and have looked at wavefront maps of different amplitude. We see that for the high amplitude wavefront map (134 nm RMS), the residual provided by the CNN is significantly smaller than the residual provided by the linear reconstructor (69 nm RMS vs 85 nm RMS), whereas, for the low amplitude wavefront map, the linear reconstructor produce slightly lower residuals (8 nm RMS vs 10 nm RMS).}

\subsection{Error by Mode}
\begin{figure}
\begin{center}
\begin{tabular}{c}
\includegraphics[width=16cm]{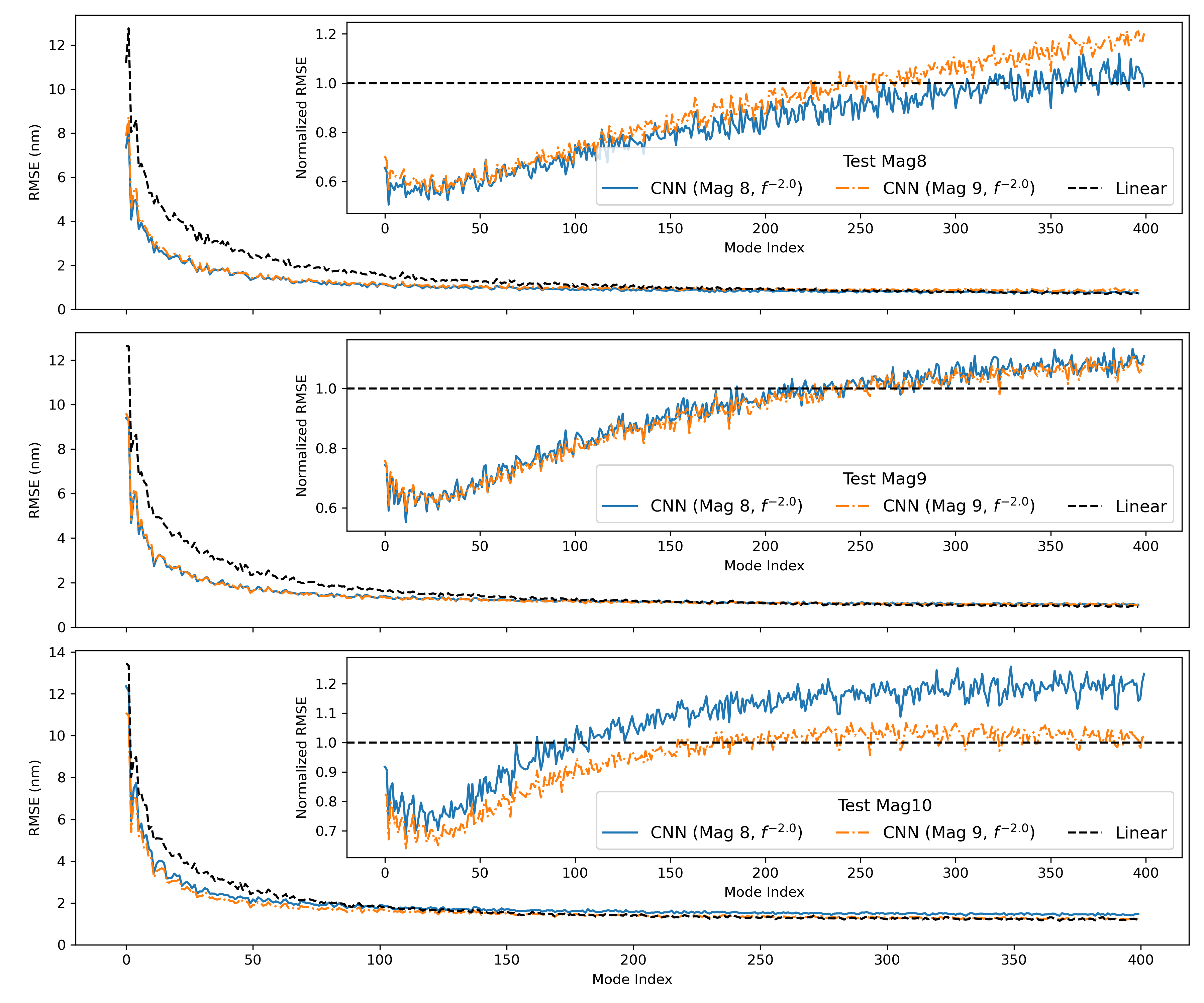}
\end{tabular}
\end{center}
\caption 
{ \label{fig:err_by_mode}
\textcolor{black}{The e}rror by mode of \textcolor{black}{the} best proposed model \textcolor{black}{on three guide stars of different magnitudes. \textcolor{black}{The m}ain panels show the CNN outperforms the linear model at lower modes.} Inset panels show \textcolor{black}{reconstruction} change relative to the linear baseline.}
\end{figure}

Our architecture \textcolor{black}{reconstructs} the first 400 modes, which are \textcolor{black}{evaluated against} the true modes for the different magnitude test sets.
By separating \textcolor{black}{reconstruction} quality as a function of mode index, Figure \ref{fig:err_by_mode} shows \textcolor{black}{that relative to the linear approximation}, our model makes the best \textcolor{black}{reconstructions} for the lowest order modes, with diminishing returns as the mode index increases\textcolor{black}{.}
Each test session consists of 2,000 frames. \textcolor{black}{The main figures show reconstruction error relative to the true coefficient.} The inset figures show the improvement in RMS wavefront error relative to the linear \textcolor{black}{reconstruction error}, which more clearly shows \textcolor{black}{that} our CNN model \textcolor{black}{outperforms the linear \textcolor{black}{reconstructor} for lower order modes}. We see that for low order modes, reduction in RMS reconstruction error of up to \textcolor{black}{4}0\% is possible, which is quite significant.

\subsection{Overall \textcolor{black}{Reconstruction} Quality}
\label{sec:overall prediction quality}
\begin{figure}
\begin{center}
\begin{tabular}{c}
\includegraphics[width=16cm]{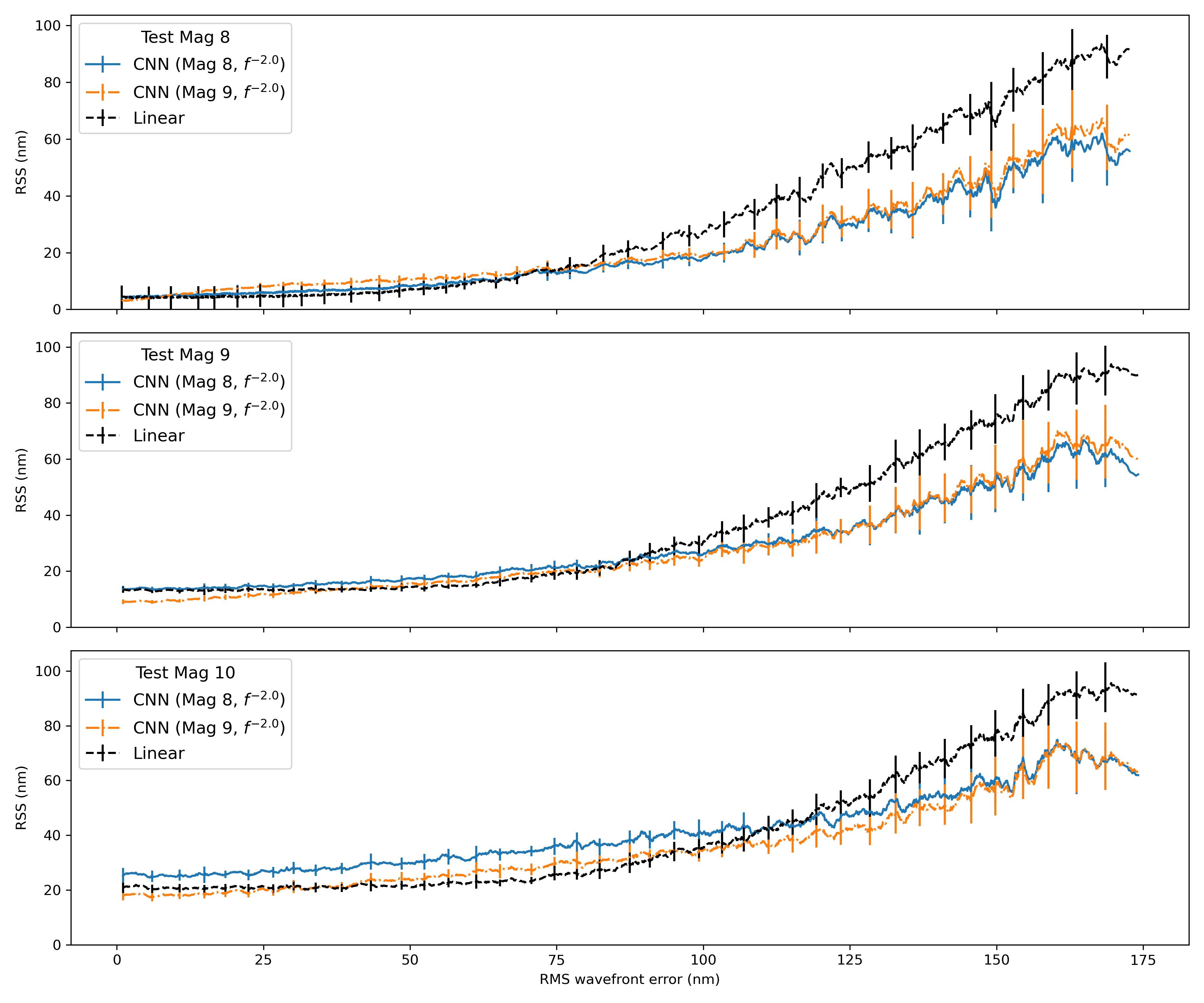}
\end{tabular}
\end{center}
\caption 
{ \label{fig:model quality}
Model \textcolor{black}{reconstruction} quality as a function of RMS for magnitudes 8, 9, and 10. Models were trained on either magnitude 8 or 9 data \textcolor{black}{following a p = -2 power law. Moving averages are plotted to help visualize the trend, with error bars showing one standard deviation for that window}.}
\end{figure}

In the results presented in the previous section, modes are examined for the entire range of 0-200 nm RMS of wavefront error. Here, we investigate the overall model \textcolor{black}{reconstruction} quality for a given RMS. Here, the RMS wavefront error is calculated using the RSS of the 1603 true modes, so it does not include the fitting error, which is why the 200 nm RMS value is not reached. Figure \ref{fig:model quality} shows \textcolor{black}{the moving average trend our} model \textcolor{black}{reconstructs} as a function of RMS. The CNN model outperforms the linear reconstruction for wavefront RMS values \textcolor{black}{above 75nm for magnitude eight sources, above 85nm for magnitude nine sources, and above 120nm for magnitude ten sources}. \textcolor{black}{These results are expected} since higher RMS values correspond to higher amplitudes and, therefore, to higher levels of non-linearities. For a wavefront error of 150 nm RMS, which is a typical residual for a PWFS-based NGS WFSs on a bright star (magnitude 8), \textcolor{black}{such as NFIRAOS (see section 1)}, the linear method results in a residual RMS wavefront error of 71 ±9 nm compared to our method, which achieved 44 ±11 nm. This reduction \textcolor{black}{is} about 40\% in RMS residual wavefront error. \textcolor{black}{For an AO system with 150 nm RMS of residual wavefront error using a linear reconstructor, the CNN would bring the residual wavefront error to $\sqrt{150^2-71^2+44^2}=139$ nm RMS. Using the Marechal approximation in H-band as in section 1, this would bring the Strehl ratio from 72.2\% to 75.6\%. When observing a point source with AO, the sensitivity, which is inversely proportional to the exposure time needed to reach a given signal-to-noise ratio, is roughly proportional to the square of the Strehl ratio\cite{Hickson2014}. Therefore, the CNN-based reconstructor could provide an increase in sensitivity of about 10\%, which is quite significant since this directly translates into a 10\% increase in observing efficiency for the telescope.}
Interestingly, the magnitude 9 \textcolor{black}{CNN} model produces reconstructions better than linear \textcolor{black}{reconstructor} in the linear region, \textcolor{black}{which is} especially visible on the magnitude 9 and 10 data. We expect this is due to the CNN model implic\textcolor{black}{i}tly applying a smoothing function that reduces the noise propagation.
Models in Figure \ref{fig:model quality} have a moving average applied with a window size of 21 to highlight the trend. Error bars show one standard deviation.

\subsubsection{Single Mode \textcolor{black}{Reconstruction}}
\begin{figure}
\begin{center}
\begin{tabular}{c}
\includegraphics[width=16cm]{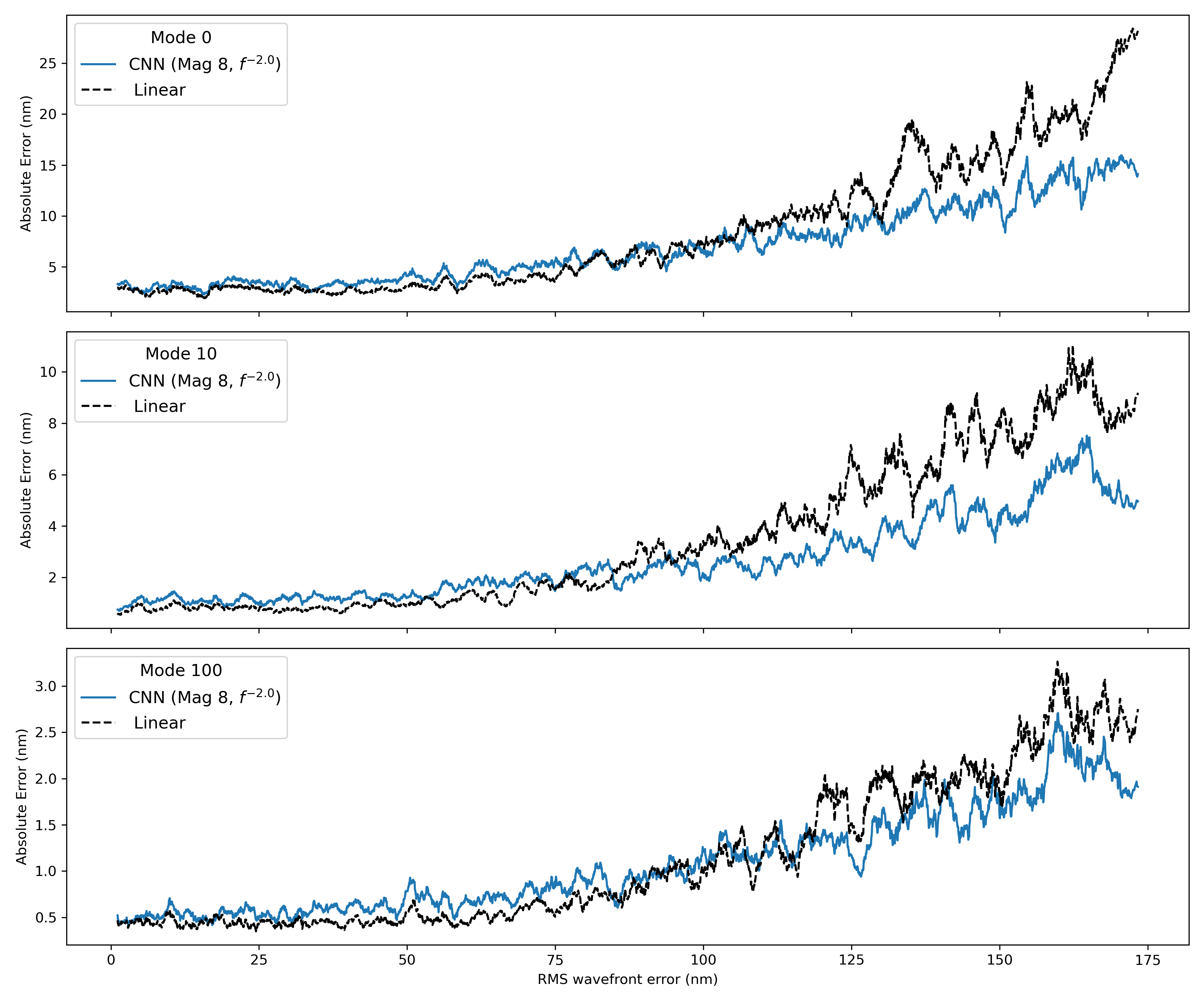}
\end{tabular}
\end{center}
\caption 
{ \label{fig:RMS quality singles}
A selection of \textcolor{black}{three} modes showing \textcolor{black}{reconstruction} quality by RMS on magnitude 8 \textcolor{black}{samples. Moving averages are plotted to help visualize the trend}.}
\end{figure}

We can inspect single modes across different values of total RMS wavefront error to better understand where the model accumulates errors. Figure \ref{fig:RMS quality singles} displays the absolute errors of modes 0 (tilt), 10, and 100 when trained and \textcolor{black}{reconstructing} magnitude 8 \textcolor{black}{samples}. The magnitude of the errors decreases with the mode index, as expected. The improvement in model \textcolor{black}{reconstruction} is most noticeable at higher RMS, and is no better than the linear model when the RMS is below 75 nm. \textcolor{black}{This is again expected since the PWFS has a linear behaviour for low RMS wavefront errors.}

\begin{figure}
\begin{center}
\begin{tabular}{c}
\includegraphics[width=16cm]{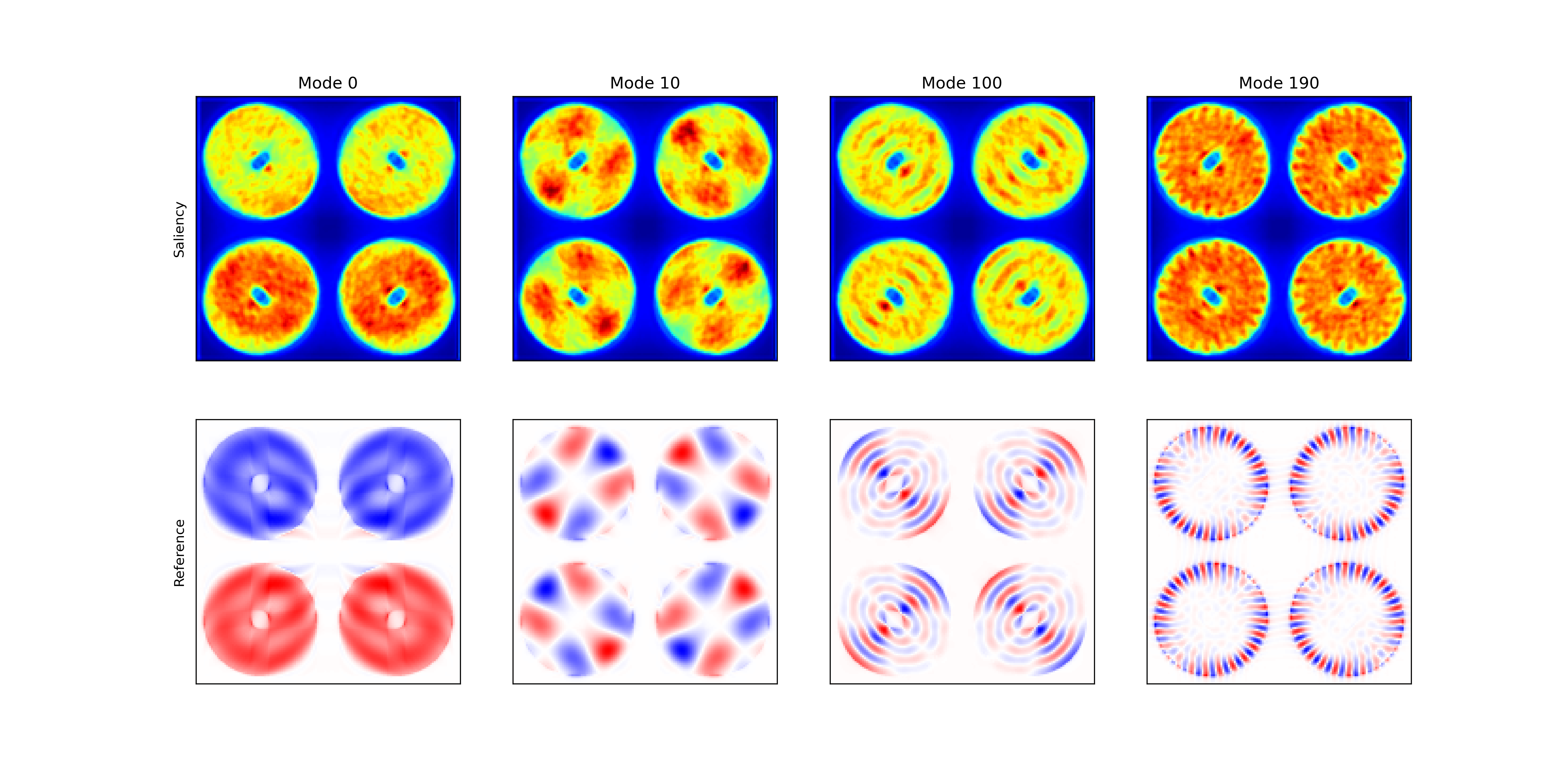}
\end{tabular}
\end{center}
\caption 
{ \label{fig:saliency map}
A selection of modes comparing saliency and idealized Karhunen—Lo\`eve modes. \textcolor{black}{The top row shows the importance of each pixel to the model when determining the modal output. The bottom row shows the corresponding Karhunen—Lo\`eve mode.}}
\end{figure}

We confirm the \textcolor{black}{CNN} gives more importance \textcolor{black}{to} the expected (physical) modes using saliency mapping. A saliency map takes a model output and determines the importance of input features used to calculate it. Multiple methods exist to combine the gradients, and we used Grad-CAM introduced by Selvaraju et al \cite{Selvaraju2017GradCAM} and implemented by the \texttt{xplique} library \cite{fel2022xplique}. The top row of Figure \ref{fig:saliency map} shows which \textcolor{black}{pixels} were important for determining the specifi\textcolor{black}{ed} mode. The bottom row provides \textcolor{black}{reference measurements of the corresponding Karhunen—Lo\`eve modes}.

\subsection{Model Robustness}
We classify model robustness in two ways; the model must be able to handle changes in atmospheric conditions and adapt to dimmer sources. Throughout the previous section, we have presented results showing how model \textcolor{black}{reconstructions} perform when we change the magnitude, and as expected, brighter sources provide more photons to reduce error in modal \textcolor{black}{reconstructions}. We now turn to robustness versus the statistical characteristics of the measured wavefronts.

\subsubsection{\textit{f}-value robustness}
\begin{figure}[h]
\begin{center}
\begin{tabular}{c}
\includegraphics[width=16cm]{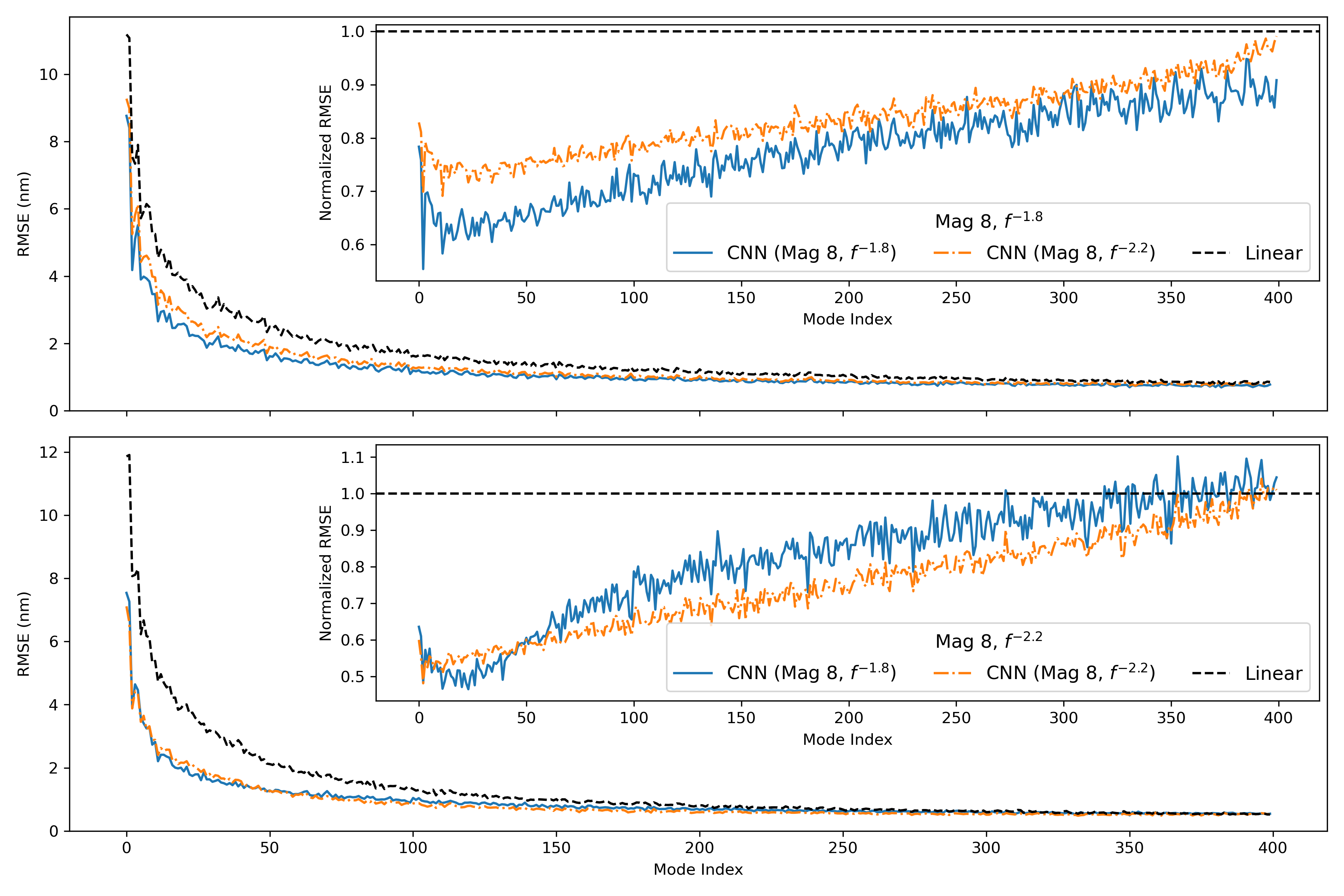}
\end{tabular}
\end{center}
\caption 
{ \label{fig:f_summary}
Two models were trained on different \textit{f}-values and testing was performed on both \textcolor{black}{sets of data. The main figures show the error against the true modal coefficient. The insert figures show error relative to the linear baseline}.}
\end{figure}

The \textit{f}-value defines a power law outlining the decreasing importance of modes. Atmospheres with higher \textit{f}-values correspond to lower frequency distortions that are captured by lower order modes. It is important that our model is not overfit\textcolor{black}{ted} to a narrow \textit{f}-value window, which would limit a real-world application. To test \textit{f}-value robustness, we prepared two models: one trained with 17,100 frames of $f^{-2.2}$ and the other trained with 17,100 frames of $f^{-1.8}$. Each model made \textcolor{black}{reconstructions} on 2,000 \textcolor{black}{samples} of both $f^{-2.2}$ and $f^{-1.8}$. The model trained on $f^{-2.2}$ acted as the baseline for the $f^{-2.2}$ testing set, and \textcolor{black}{reconstructions} made by the model trained on $f^{-1.8}$ showed how the \textcolor{black}{reconstruction} quality drops with a change in \textit{f}-value. This was repeated for the $f^{-1.8}$ model acting as the baseline. Figure \ref{fig:f_summary} shows the drop in model performance caused by changing the \textit{f}-value during testing. Both models performed significantly better than the linear model, though not as well as the one trained directly on the corresponding \textit{f}-value data. Interestingly, the loss in \textcolor{black}{reconstruction} quality was anisotropic with the normalized RMSE of both models having different characteristics: $f^{-1.8}$ made better \textcolor{black}{reconstructions} for a small window of modes on the $f^{-2.2}$ test set, while the $f^{-2.2}$ model performed uniformly worse on the $f^{-1.8}$ test set. These observations suggest the models were learning filters to prioritize modes more common in their training data.

\begin{figure}[h]
\begin{center}
\begin{tabular}{c}
\includegraphics[width=16cm]{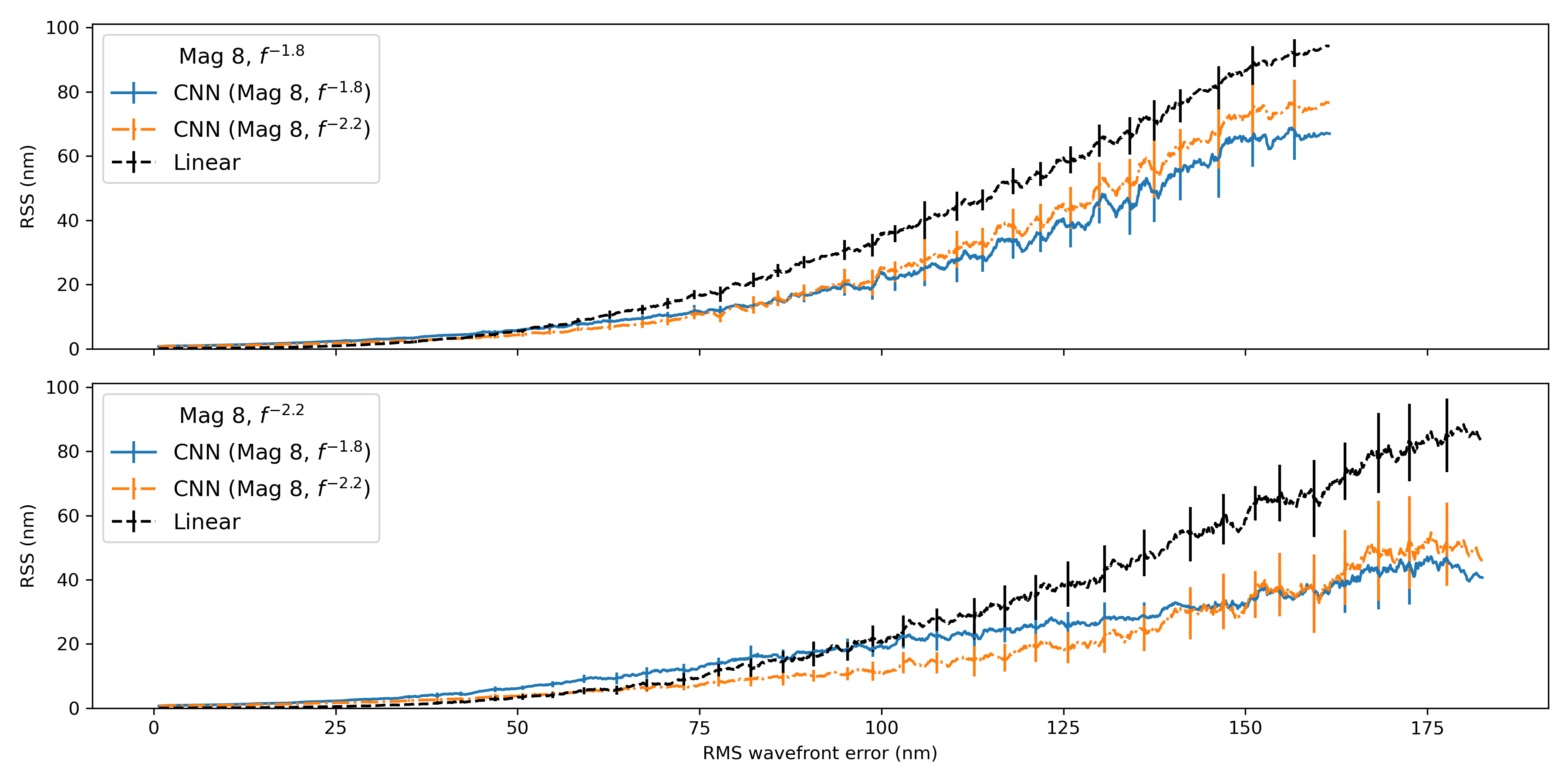}
\end{tabular}
\end{center}
\caption 
{ \label{fig:f_summary_rms}
The error by RMS of two models trained on different \textit{f}-values. The upper panel shows models \textcolor{black}{reconstructing} $f^{-1.8}$ \textcolor{black}{samples} and the lower \textcolor{black}{reconstructing} $f^{-1.8}$ \textcolor{black}{samples. Moving averages are plotted to help visualize the trend. Error bars are one standard deviation for that window}.}
\end{figure}

\textcolor{black}{These two models were then evaluated by looking at frame quality as a function of RMS in the same way as in section \ref{sec:overall prediction quality}. In Figure \ref{fig:f_summary_rms} we see both models outperforming the linear model for higher RMS frames (above 100nm). Here we can more clearly see that the model trained on $f^{-1.8}$ atmospheric conditions does not transfer to $f^{-2.2}$ atmospheric conditions as well as the inverse.}

\section{Conclusion and Future Work}
\label{sec:discussion}

\textcolor{black}{In this work, we demonstrated that a neural network is able to take into account non-linearities in the measurement process of an AO PWFS and make considerable improvements over a linear reconstructor in delivered image quality}. These improvements are most noticeable in the low order modes, and for RMS errors above 75 nm, which is typical for astronomical AO systems. When measuring a 150 nm RMS wavefront residual, the \textcolor{black}{portion due to} reconstruction errors is found to be reduced by about 40\%, which \textcolor{black}{results in a gain of more than 3 points in Strehl ratio at H-band (1.65 {\textmu}m). When observing a point source, this gain could translate to an increase in observing efficiency of up to 10\%, which is quite significant.}. 

We see diminished returns on fainter sources where WFS noise dominates, but still potential improvements \textcolor{black}{are observed in higher ($>$ 125nm) RMS conditions. A result of focusing on the overall \textcolor{black}{reconstruction} means that \textcolor{black}{the} quality of a mode \textcolor{black}{reconstruction} decreases with index, which can be explained by our model using convolutional filters to identify lower frequency patterns. Increasing the number of filters - especially at the lower layers - should improve the \textcolor{black}{reconstruction} quality, but this was not found in practice. The cause for the difficulty \textcolor{black}{reconstructing} higher order modes with our CNN architecture is unclear but may be related to these modes concentrating power towards the edge of the pupils. Why exactly that would be a problem and how to mitigate it is future work.}

A major goal of this work is to test the robustness of a fixed model to \textcolor{black}{reconstructions} outside of its \textcolor{black}{training} range. This occur\textcolor{black}{s} in two dimensions: changes in magnitude, and changes in $f$-value. The improvements for low order mode \textcolor{black}{reconstructions} remained for magnitude 10 data when the model was trained on magnitude 8 data. When our model was trained on magnitude 9 data, this robustness was further realized, suggesting that training on more varied samples could further improve the final model.
The other axis of robustness, $f$-value change, was evaluated by training models on different power-law distributions of residual wavefront error. These models were then evaluated on the opposing dataset, where we found they were susceptible to changes in this f-value. This change is not unexpected as the convolutional layers in the model would be more tuned for higher or lower frequency patterns depending on the training set.

\textcolor{black}{This paper is focused on proof-of-concept through numerical simulations, and, at this point, we have not thought in details on how this approach could be implemented on a real system. For the training, one could simply use simulated wavefront maps and simulated PWFS measurements, making sure that the parameters of the target AO systems are correctly captured in the model. For a more realistic training, based on physical measurements, one could imagine installing a high precision wavefront sensor at the output of the AO system, and a wavefront generation device, such as a deformable mirror or a Spatial Light Modulator or a phase screen at the input, the latter providing sample wavefront maps that the PWFS could measure while the former would provide the truth measurement. Since training can be performed at a relatively slow speed if necessary, commercial devices could be used. The other challenge is to run the ML-reconstructor in real time, with a latency low enough to track the atmospheric turbulence (typically $<$ 1 ms). This implementation would replace the matrix-vector multiply used to implement the linear reconstructor in conventional AO systems and probably requires a combination of architecture optimization and advanced computation hardware.}

\textcolor{black}{Other future work includes training on more varied datasets to improve robustness, and investigating the potential benefits of using different CNN models for different observing conditions.}

% \disclosures 

\section* {Code, Data, and Materials Availability} 
Code is available upon request.
%As relevant, declare the availability of computer software code, data, and/or materials used in the research results reported in the manuscript. Provide specific access information or restrictions for code, data, and materials (i.e., links to repository access addresses, and/or guidance on commercial or public access). Note: reporting in this section is required for the \textit{Journal of Biomedical Optics} and \textit{Neurophotonics}. 

%%%%% References %%%%%

\bibliography{report}   % bibliography data in report.bib
\bibliographystyle{spiejour}   % makes bibtex use spiejour.bst

%%%%% Biographies of authors %%%%%

\vspace{2ex}\noindent\textbf{Finn Archinuk} is a Masters student at the University of Victoria in the Computational Biology Research and Analytics (COBRA) lab. He received his BS degree in Microbiology at the University of Victoria. He has a varied publication resume including secondary metabolism in Poplar and galaxy quantization. He briefly worked as a machine learning researcher at National Resource Council Canada, where this work originates.

%\vspace{1ex}
%\noindent Biographies and photographs of the other authors are not available.

%https://www.spiedigitallibrary.org/journals/journal-of-astronomical-telescopes-instruments-and-systems/author-guidelines

\end{spacing}
\end{document}